\documentstyle[prd,aps,graphicx]{revtex}
\newcommand{\beq}{\begin{equation}}
\newcommand{\enq}{\end{equation}}
\newcommand{\Sch}{Schwarzschild}
\newcommand{\beqn}{\begin{eqnarray}}
\newcommand{\enqn}{\end{eqnarray}}
\newcommand{\pa}{\partial}
\draft
\begin{document}

\twocolumn[\hsize\textwidth\columnwidth\hsize\csname
@twocolumnfalse\endcsname

\title{Initial Data for Black Holes and Black Strings in 5d.}
\author{Evgeny Sorkin and Tsvi Piran}
\address{ Racah Institute of
  Physics, The Hebrew University, Jerusalem, 91904, Israel.}
\maketitle

\begin{abstract}
  We explore time-symmetric hypersurfaces containing apparent horizons
  of black objects in a 5d spacetime with one coordinate compactified
  on a circle. We find  a phase transition within the family of such
  hypersurfaces: the horizon has different topology for different
  parameters. The topology varies from $S^3$ to $S^2 \times S^1$. This
  phase transition is discontinuous -- the topology of the horizon
  changes abruptly.  We explore the behavior around the critical point
   and present a possible phase diagram.
\end{abstract}

\vskip1pc]

Several black objects solutions exist in spacetimes with
dimensionality greater then four containing compact dimensions.  Among
these solutions are black strings (BS)  and black holes (BH)-- two
black objects with distinct horizon topology. However no general
analytic solution is known for a black object in a compactified
space-time in more then 4d spacetime.

Gregory and Laflamme (GL) \cite{GL1,GL2} discovered that a uniform
black string, a product of a \Sch\ solution with a line, develops a
dynamical instability if a compactification radius is 'too large'.
They postulated the existence of a new branch of non-uniform black
string solutions.  The endpoint of an unstable uniform string was
expected to be a BH.  However, Horowitz and Maeda
\cite{HM1,HM2} argued that the GL instability cannot cause a uniform
BS to decay to a BH, at least in a finite affine
time. This is because of the 'no tear' property of horizons.  They
conjectured that the endstate of such a decay would be rather a
non-uniform BS. Non-uniform BS solutions connected
to the GL point were constructed perturbativly by Gubser
\cite{gubser}. Wiseman \cite{wiseman1} found non-uniform
BS solutions using fully non-linear numerical calculations.
However, Wiseman stresses that these solutions are most likely
unstable.

We consider here  the five dimensional spacetime  with one compact
dimension of length $2L$.  We denote the coordinate along the compact
dimension by $z$. A neutral black object has a characteristic
dimensionsfull parameter, its asymptotic 4d mass, $M$. It is
convenient to describe the system by a single dimensionless parameter,
$\zeta \equiv G_4 M / (2 L)$, where the effective Newton constant is
$G_4=G_5/2L$. If $\zeta \ll 1$ the black object is expected to
resemble a 5d \Sch\ BH since local measurements cannot probe
the compactness of the $z$ direction.  The horizon of this BH is
therefore expected to have an $S^3$  topology.  At the opposite
end,  $\zeta \gg 1$, one expects that the horizon would extend
over the compact dimensions wrapping it completely. The
horizon's topology of this object, $S^2 \times S^1$, suggests the
name 'BS '.  An uniform BS is described by a 4d
\Sch \ solution times a circle. A non-uniform BS is characterized
by a non-trivial z-dependence along the circle.  An attempt,
based on a thermodynamical reasoning, to construct the  phase diagram
that would include consistently different phases of solutions was
recently undertaken by Kol \cite{kol}.

In this letter we make a step toward the self-consistent study of
black objects in compact 5d background. We solve numerically the
initial-value problem for a moment of time-symmetry for spacetimes
containing black objects.  We determine the {\it apparent} horizon of
the existing black objects in these solutions and we follow the
transition from spherical to cylindrical topology in a family of
solutions with a different dimensionless parameter $\zeta$. Although
the configurations we find are not 5d static solutions, these
solutions, give us some insight on the behavior of the black objects
in a compactified background. A similar approach was previously
applied \cite{shibata} to explore the time-symmetric BH problem in the
brane world scenario.

We consider a time-symmetric slice through the black object's
space-time.  To generate the black object solution we consider a
configuration with an artificial matter around the origin.  This
matter gives rise to a non-flat spatial metric along the slice.  We
solve numerically for this metric. Then we solve for the {\it
  apparent} horizon of the black object. The method allows us to
determine the spatial geometry of the black object. Since we solve
along a time-symmetric slice, this spatial geometry constitutes the
initial data for subsequent dynamical time evolution of this black
object's geometry. Limiting ourselves only to the determination of the
initial data we find explicitly time-symmetric solutions with a
distinct horizon topology and construct the phase diagram for a family
of such solutions parametrized by $\zeta$.  We show that there is a
phase transition within this family. We identify the critical
$\zeta_c$ and show explicitly that if $\zeta_c$ is reached from below
a corresponding BH solutions becomes deformed and the horizon
has a cigar-like shape.  If $\zeta_c$ is approached from above, the
corresponding BS solution becomes more and more non-uniform.
However, this cannot be considered as evidence for existence of a
non-uniform static BS solutions.  This initial data may
relax to a uniform string during its dynamical evolution.  We find
that the transition between both topologies is not smooth: the
spherical-like horizon jumps suddenly to become a cylindrical-like
horizon. Put differently, neither the BS pinches nor the
BH has its south and north poles touching each other.

The method that we describe here enables us to study properties of
black objects with different apparent horizon's topology in a single
consistent numerical scheme.  Our method does not require a
prescription of the topology of the horizon.  The topology is not
predetermined but rather it is a derived result of numerics.
Usually, to obtain a {\it static} black object solution of the 5d
Einstein equations one would had to specify the topology of the
horizon. In other words there is no a single numerical scheme that can
find black objects with distinct horizon
topology.

Let us  consider a time-symmetric, spacelike hypersurface $\Sigma_t$ with
a vanishing extrinsic curvature. The  Hamiltonian constraint reads
\beq
\label{hamcon}
R^{(4)} =16 \pi G_5 T_{\mu\nu} t^\mu t^\nu \ ,
\enq
where $t^\mu$ is the unit normal to $\Sigma_t$ and $ T_{\mu\nu}$ is
the 5d stress-energy tensor. From now on we work in units where
$c=G_5=1$.  The momentum constraint is trivially satisfied provided
that the matter is static and the extrinsic curvature of this slice
vanishes.

We choose the metric on $\Sigma_t$ to be conformally flat
\beq
\label{metric}
dl^2= \psi^2 ( dr^2 +  dz^2 + r^2 d\Omega_2^2 ) \ .
\enq
This ansatz has been adopted for simplicity. We didn't expect that the
static solution will be conformally flat. However, we expect that the
trend we discuss is insensitive to this assumption. The method can be easily
generalized to non-conformal choices. More complicated metrics will be
considered elsewhere.

With this choice of the metric the constraint equation takes the
form %
\beqn
\label{equation1}
\nabla^2 \psi +\rho \psi^3 =0 \ , \rm{where} \ \nabla^2\equiv {1\over
  r^2} \pa_r\left(r^2 \pa_r \right) + \pa_{zz} \ .  \enqn Here we have
defined $\rho \equiv T_{\mu\nu} t^\mu t^\nu >0,$ the physical energy
density. This $\rho$ is the matter density, distributed around the
origin. In practice, we pick a localized matter distribution that
occupies some finite region around the origin.  This artificial matter
gives rise to a non-trivial $\psi$ in a way that does not involve
specifying an inner boundary of a black object. For sufficiently
concentrated mass an apparent horizon appears. For lower values of
$\rho$ the solution describes a momentary static star.

To solve the elliptic Eqn (\ref{equation1}) we have to specify
boundary conditions.  These conditions are: (i) At the equatorial
plane, $z=0$, we have a reflection b.c.  $\partial_z \psi = 0$; (ii)
At the symmetry axis, $r=0$, we have also a reflection symmetry:
$\partial_r \psi = 0$; (iii) Periodicity in $z$ implies also
reflection symmetry at $z= L$: $\partial_z \psi = 0$; (iv) At
$r\rightarrow \infty$ we set $\psi \rightarrow 1+G_4 M/ r$, which is
consistent with the 4d asymptotic behavior\footnote{The asymptotic
  metric is expected to take the form of a 4d \Sch \ times a line.
  This metric generally cannot be brought to the conformally flat form
  (\ref{metric}). However, asymptotically, as $r\rightarrow \infty$, a
  transformation is possible at the leading order. The asymptotic
  behavior of the conformal factor would become $\psi \rightarrow
  1+G_4 M/ r$. The length of the circle tends asymptotically to $2L$ so we
  can read the asymptotic 4d mass in the units of the effective Newton
  constant $G_4 = G_5/2L$.}. This condition is implemented by
rewriting it as $ \pa_r (r \psi)=1$, eliminating the need to specify
$M$. Instead, $M$ is determined from the numerical solution.

There is a wide literature about equations like (\ref{equation1}), see
\cite{York1} and references therein. This equation is ill posed and it
does not have a unique solution. In order to bring the equation to a
well posed form we rescale the matter density $\rho\rightarrow
\tilde{\rho} \psi^{-3-s}$, with some $s>0$ and  nonphysical density $
\tilde{\rho}$, see e.g.  \cite{York1}. The resulting equation
\beqn
\label{equation}
\nabla^2 \psi +\tilde{\rho}\psi^{-s} =0 \
\enqn
is well posed and has a unique solution. The rescalling of the matter
density is unimportant as we are interested only in the external
vacuum part.  The  solution for $\psi$ is found using
relaxation.

We obtain a sequence of momentary time-symmetric solutions by fixing
the density of the artificial matter and its location as
$\tilde{\rho}=10^6 \Theta(0.5-r) \Theta(0.5-z) $, and varying
continuously the length of the compact asymptotic circle, $2L$.  The
figures below are obtained for this  source.  We checked that
our results are not affected by the specific choice of the source.
Taking the smooth distribution $\tilde{\rho}=10^6
\exp(-r^2/\sigma_r^2)\left[\exp(-z^2/\sigma_z^2)+\exp(-(z-2L)^2/\sigma_z^2)\right]$
with various $\sigma_r$ and $\sigma_z$ we found the same overall
picture. The values of $\zeta_c$ varied, as expected, depending on the
source. The variations in $\zeta_c$ are $\zeta_c =.98-1.8$.  Moreover,
taking $\sigma_r\simeq 0.5 , \sigma_z \gg L$, i.e.  practically
cylindrical source, we were able to reproduce the uniform BS
solution. The position of the horizon in this solution (see Eq.
(\ref{0expansion}) and the subsequent discussion) was determined to
within $0.1\%$. This provides an independed check on the overall
accuracy of our calculation in addition to other standard tests of
convergence, errors scaling etc.

To solve the equation for $\psi$ we used grids, covering
the domain $0<z<L$, and $0<r<R_{cut}$, with typical grid spacings of
$\Delta r \sim 0.02$ and $\Delta z \sim 0.01$. The value of $R_{cut}$,
where the grid was cutoff and the asymptotic b.c. (iv) was
implemented, has been taken as $R_{cut}=5,10$ and $20$.  We checked
that the results are insensitive to variation of $R_{cut}$, provided
that $R_{cut} > 5$.

\begin{figure}[b!]
\centering
\noindent
\includegraphics[width=8.8cm,height=8cm]{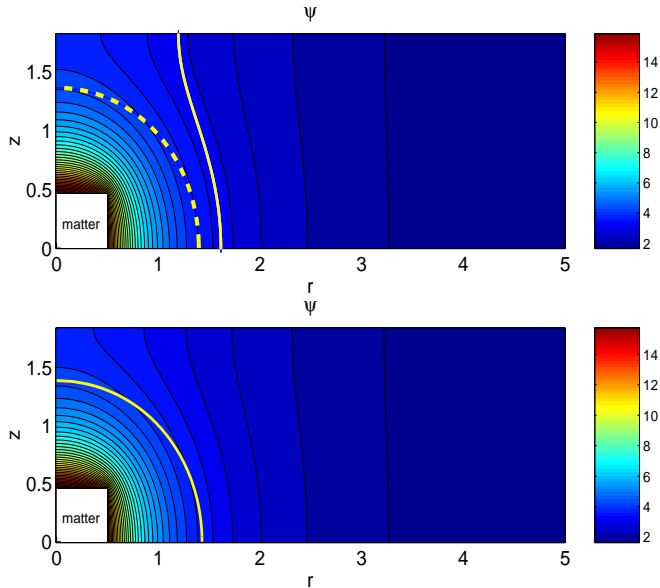}
\caption{Contours of $\psi$. Upper panel: $\zeta \simeq 1.85$ and
  there is a non-uniform BS.  Bottom panel: $\zeta \simeq
  1.7$ and there is a BH. In both plots one observes how the
  deformation of the contour lines fades asymptotically.  We plot also
  the apparent horizons in both cases. These are designated by thick
  curves. In the upper plot that corresponds to the BS phase
  there are two horizons. The inner spherical apparent horizon,
  designated by the dotted thick curve, and the outer cylindrical
  horizon, designated by the solid curve.}
\label{fig:cont}
\end{figure}

To envisage the spatial metric around the black objects in
Fig.\ref{fig:cont} we plot the contours of $\psi$ in two cases that
correspond to a BH solution and a non-uniform BS
solution. The matter is located near the origin and is encircled by
the horizon in either case. The geometry outside the apparent horizon
 has an axisymmetric structure and it becomes asymptotically
flat. The $\psi$ contour lines are spherical near the origin and
become cylindrical as $r$ increases.

Once we obtain $\psi$ we determine the existence of the apparent
horizon.  An apparent horizon is defined by a zero-expansion of the
null rays generating the horizon\cite{Hawk}. For the time-symmetric
hypersurface $\Sigma_t$ this condition can be written as
\beq
\label{0expansion}
 \nabla^{(4)}_\mu n^\mu=0 \ ,
\enq
where $n^\mu$ is the the unit normal to the apparent horizon.

To simplify the treatment, we distinguish between two different
topologies for the horizon.

(1) When the horizon has the topology $S^2 \times S^1$ we choose
cylindrical coordinates $(r,z)$ and   we solve for a curve $r=h(z)$.

(2) When the horizon has the topology of $S^3$ it is convenient to
transform to spherical coordinates $R,\chi$ defined by $r=R
\sin(\chi), z=R \cos(\chi)$ The horizon in the $R,\chi$ plane is
given by a curve $R=h(\chi)$.

The unit normal to
the curve that defines the horizon is
\beq
n^\mu=(C,-C h')\ .
\enq
The parameter $C(r,z)$ could be read from the normalization condition
$n^\mu n_\mu =1$.  In both cases we solve numerically equation
(\ref{0expansion}) to obtain the position of the apparent horizon.

  A useful qualitative parameter employed as a measure of the
 non-uniformity of a BS \cite{gubser} is $\lambda \equiv 1/2
 (r_{max}/r_{min}-1)$ where $r_{min}$ and $r_{max}$ are the minimal
 and the maximal 4d \Sch \ radii of the apparent horizon. For a
 uniform string $\lambda=0$. In the BH phase $\lambda=\infty$.
 Therefore, for a BH we define {\it another} parameter $\lambda'\equiv
 R_{max}/R_{min}-1$, where $R_{max}$ and $R_{min}$ are the 5d \Sch \
 radii of the horizon. This parameter gives an idea of deformation of
 the BH's horizon.

\begin{figure}[b!]
\centering
\noindent
\includegraphics[width=8.8cm,height=8cm]{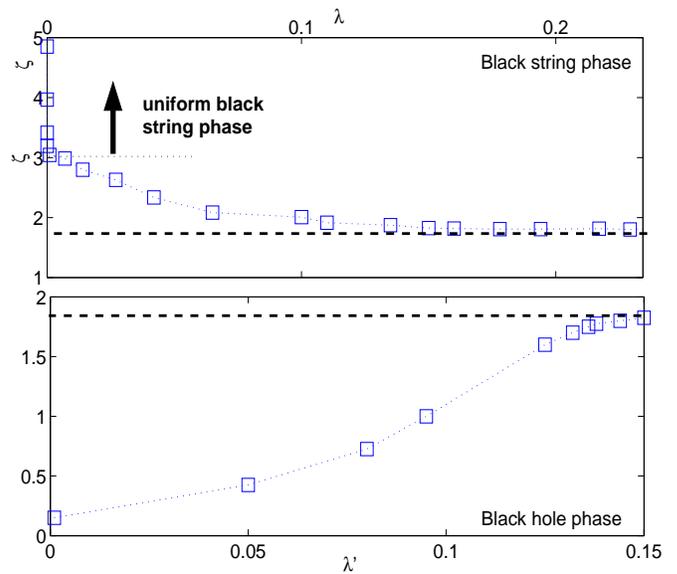}
\vspace{0.1cm}
\caption{The path in the configuration space of the numerical
  solutions. The BH (BS) phase is displayed in the bottom (top) panel.
  In both
  panels the horizontal dashed line that is located at $\zeta_c\simeq
  1.78$ designates the $BS\rightarrow BH$ transition
  point. Above this line there is BS phase, while below the
  line there is the BH phase. The BS becomes uniform
  at $\zeta \simeq 3.0$ and the BH becomes spherical at
  $\zeta \simeq 0.15.$ }
\label{fig:pd}
\end{figure}
 We find that there are two topologically distinct apparent horizon
 solutions. At small $\zeta$ the topology of the horizon is $S^3$ and
 the horizon is close to be exactly spherical.  When $\zeta$ increases
 we see that the horizon begins to deform, deviating from a spherical
 shape but still remaining topologically 3-sphere. At a certain value,
 $\zeta_m \simeq 1.78$, a phase transition takes place -- the topology
 of the apparent horizon changes form $S^3$ to $S^2\times S^1$. In
 fact there are two apparent horizons in the BS phase. The outer
 horizon, has a cylindrical topology, while the inner one has a
 spherical topology.  In Fig.\ref{fig:pd} we plot the sequence of
 solutions parametrized by $\zeta$. In this figure there are finite
 values of $\zeta$ when $\lambda=0$ or $\lambda'=0$. In fact for these
 $\zeta$ the deformation of the horizon becomes so small that cannot
 be resolved by our numerics and we put the corresponding lambdas to
 zero.

Another interesting result is the measure of geometrical deformation
of the horizons. In the BS and the BH phases this
measure is supplied by $\lambda$ and $\lambda'$ respectively. The
non-uniformity of the BS is displayed in the upper  panel
of Fig. \ref{fig:pd}. Near the critical point  the most non-uniform
string has $\lambda \simeq 0.22$. The non-uniformity disappear as
$\zeta$ increases. For $\zeta \geq 3.0$ the BS becomes a
uniform BS. The deformation of the horizon in the BH
phase could be seen in the bottom panel of the same Figure. The
maximal radius, $R_{max}$,  always occurs at the axis, $r=0$. The
BH becomes more and more oblate and  stretched along the symmetry
axis, as we  approach the critical point. The most deformed
BH has $\lambda' \simeq 0.15$ just before the transition.

\begin{figure}[b!]
\centering
\noindent
\includegraphics[width=8.0cm,height=4cm ]{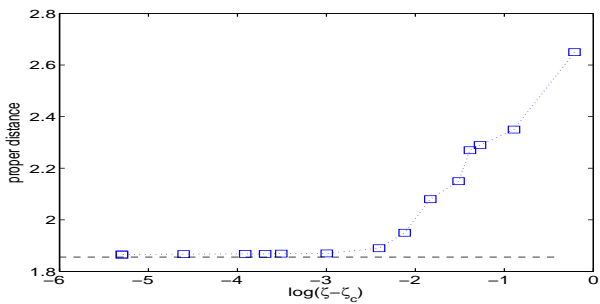}
\vspace{0.1cm}
\caption{ The proper distance from the BH to the
  reflection plane $z=L$ does not decrease to zero but tends to  a
  finite value as we approach the transition point. }
\label{fig:ell}
\end{figure}

The phase transition is discontinuous. At the critical
value of $\zeta_c$ the spherical horizon jumps suddenly to become
a cylindrical. To get insight on the behavior of the phase
transition, in the BH phase we have computed the proper
distance along the $r=0$ axis from the the BH horizon at the
axis to $z=L$:
\beq
\label{prop}
\ell=\int^L_{r^{(5)}} {\psi(r=0,z) dz} \ . \enq
This distance decreases as we increase $\zeta$. One could expect that
as horizon grows and as $\ell \rightarrow 0$ the north pole of the
BH would tend towards its south pole and they will touch.
However, we find that as $\zeta \rightarrow \zeta_c$ $\ell$ reaches a
finite value.  We plot the behavior of $\ell$ as a function of
$\zeta-\zeta_c$ in Fig. \ref{fig:ell}. One observes that $\ell$ tends
to a positive constant just before the transition.

The initial data that we have constructed here is analogous
to the Misner initial data \cite{misner} for a family of
momentary static two equal mass BHs in 4d GR. Misner choose a
sequence  of conformally flat metrics with the conformal factor
parametrized by a certain parameter $\mu$ that is related to the mass
of the BHs and their proper mutual separation. As $\mu$ varies
the shape of the initial apparent horizons varies. If the BHs
are close enough, that occurs for small $\mu$, a new apparent horizon
suddenly appears, surrounding both BHs on the initial
hypersurface. In other words, there is a critical $\mu_0$ that divides
two distinct possibilities for the topology of the apparent horizon on
the initial slice. Just by looking at this initial data sequence one
has an indication that the event horizons for two BHs will
merge and form a distorted BH during an actual evolution. The
value of $\mu$ when this merge occurs, generally would not coincide
with the theoretical, $\mu_0$. In fact, the numerical evolution
\cite{cbh1,cbh2} of Misner initial data shows that the qualitative
picture obtained for the sequence of the initial data is correct. The
actual critical value of $\mu$ does not coincide with $\mu_0$, they
are, however, not that different from each other.

Here we have an infinite array of BHs that are approaching
each other simultaneously. We have shown that there is a family of
initial data parametrized by $\zeta$.  When the separate BHs
in the array are getting closer they become distorted form the
spherical shape.  At a critical value $\zeta_c$ the separate horizons
are engulfed suddenly by a single cylindrical-like horizon.  The
effective cylindrical horizon after the transition is non-uniform.

It
is important to stress that the sudden jumps of the apparent horizon
topology are different from the first order transition that can happen
between different phases of static solutions, discussed in
\cite{gubser,kol}. This is because the apparent horizon that we
discuss here is not casual as it is defined only locally. The event
horizon is a global concept and it is expected to be larger than the
apparent horizon. Only for static solutions both horizons coincide.
Since our solution is not static the jumps of  the apparent horizon  cannot
exclude the possibility of a smooth transition between the BHs
and the BSs event horizons, discussed in \cite{ObHer}.

The
concrete numerical value of $\zeta_c$ isn't important as it is just a
number characteristic to the specific initial data sequence. However we
expect that the qualitative behavior would be similar in the static
solutions as well. We believe that a dynamical evolution of our
initial data would confirm this qualitative picture and will yield
actual critical value, $\zeta_c$.

{\bf Acknowledgment} We  thank Barak Kol for useful conversations
and an ISF grant for support.


\end{document}